\documentclass[%
reprint,
superscriptaddress,
amsmath,amssymb,
aps,longbibliography,
]{revtex4-1}

\usepackage{graphicx}%
\usepackage{verbatim}
\usepackage[version=4]{mhchem}
\usepackage{xcolor}
\usepackage{physics}
\usepackage[normalem]{ulem}
\usepackage{pdfpages}
\usepackage{pgffor}

\makeatletter
\AtBeginDocument{\let\LS@rot\@undefined}
\makeatother

\ifdefined\COSMOMATHS
\else
\newcommand{\COSMOMATHS}{}

\newcommand{\mbf}[1]{\ensuremath{\mathbf{#1}}}

\newcommand{\D}[1]{\operatorname{d}{\!#1}\,}

\fi %

\ifdefined\DIRACREP
\else
\newcommand{\DIRACREP}{}

\usepackage{physics}
\usepackage{xparse}
\ifdefined\COSMOMATHS
\else
\newcommand{\COSMOMATHS}{}

\newcommand{\mbf}[1]{\ensuremath{\mathbf{#1}}}

\newcommand{\D}[1]{\operatorname{d}{\!#1}\,}

\fi %

\NewDocumentCommand{\rep}{s d<| d|>}{%
\IfBooleanTF{#1}{
   \IfValueTF{#2}{
       \IfValueTF{#3}{\braket{#2}{#3}}{\bra{#2}}
       }{
       \IfValueTF{#3}{\ket{#3}}{}
       }
   }{
   \IfValueTF{#2}{
       \IfValueTF{#3}{\braket*{#2}{#3}}{\bra*{#2}}
       }{
       \IfValueTF{#3}{\ket*{#3}}{}
       }
   }
}

\NewDocumentCommand{\rbra}{sm}{\IfBooleanTF{#1}{\rep*<#2|}{\rep<#2|}}
\NewDocumentCommand{\rket}{sm}{\IfBooleanTF{#1}{\rep*|#2>}{\rep|#2>}}
\NewDocumentCommand{\rbraket}{smom}{
    \IfBooleanTF{#1}{
        \IfNoValueTF{#3}{\rep*<#2||#4>}{\rep*<#2|#3\rep*|#4>}
    }{
        \IfNoValueTF{#3}{\rep<#2||#4>}{\rep<#2|#3\rep|#4>}
    }
}

\NewDocumentCommand{\cg}{m m m}{\rep<#1; #2||#3>}

\NewDocumentCommand{\field}{o m e{_} e{^} o e{_} e{^}}{
\IfValueTF{#5}{\overline{
  #2\IfValueT{#3}{_#3}\IfValueT{#4}{^{\otimes #4}} %
  \otimes
  #5\IfValueT{#6}{_#6}\IfValueT{#7}{^{\otimes #7}} %
  \IfValueT{#1}{;#1}
}}{
  \IfValueTF{#4}{\overline{
     #2\IfValueT{#3}{_#3}\IfValueT{#4}{^{\otimes #4}}
     \IfValueT{#1}{;#1}
  }}
  {#2\IfValueT{#3}{_#3}}
}
}

\NewDocumentCommand{\frho}{o e{_} e{^}}{
\field[#1]{\rho}_{#2}^{#3}
}

\newcommand{\br}{\mbf{r}}
\newcommand{\bx}{\mbf{x}}

\newcommand{\e}{a}  %

\NewDocumentCommand{\ex}{e_}{
\IfValueTF{#1}{\e_{#1}\bx_{#1}}{\e\bx}
}  %

\NewDocumentCommand{\lm}{e_}{
\IfValueTF{#1}{l_{#1}m_{#1}}{lm}
}
\NewDocumentCommand{\nlm}{e_}{
\IfValueTF{#1}{n_{#1}\lm_{#1}}{n\lm}
}
\NewDocumentCommand{\enlm}{e_}{
\IfValueTF{#1}{\e_{#1}\nlm_{#1}}{\e\nlm}
}
\NewDocumentCommand{\en}{e_}{
\IfValueTF{#1}{\e_{#1}n_{#1}}{\e n}
}
\NewDocumentCommand{\nlk}{e_}{
\IfValueTF{#1}{n_{#1}l_{#1}k_{#1}}{nlk}
}
\NewDocumentCommand{\enlk}{e_}{
\IfValueTF{#1}{\e_{#1}\nlk_{#1}}{\e\nlk}
}
\NewDocumentCommand{\enl}{e_}{
\IfValueTF{#1}{\en_{#1}l_#1}{\en l}
}

\NewDocumentCommand{\nnl}{s}{
\IfBooleanTF{#1}{n_1 n_2 l}{n_1; n_2; l}
}
\NewDocumentCommand{\ennl}{s}{
\IfBooleanTF{#1}{\en_1 \en_2 l}{\en_1; \en_2; l}
}

\NewDocumentCommand{\gslm}{s}{
\IfBooleanTF{#1}{\sigma\lambda\mu}{\sigma;\lambda\mu}
}

\newcommand{\rcut}[0]{{r_\text{cut}} }

\fi %

\ifdefined\COSMOMODELS
\else
\newcommand{\COSMOMODELS}{}
\usepackage{bm,upgreek}

\newcommand{\krn}[0]{\operatorname{k}}

\fi %

\begin{document}

\title{Wigner kernels: body-ordered equivariant machine learning without a basis}

\author{Filippo Bigi}
\affiliation{Laboratory of Computational Science and Modeling, Institut des Mat\'eriaux, \'Ecole Polytechnique F\'ed\'erale de Lausanne, 1015 Lausanne, Switzerland}
\thanks{These two authors contributed equally}

\author{Sergey N. Pozdnyakov}
\affiliation{Laboratory of Computational Science and Modeling, Institut des Mat\'eriaux, \'Ecole Polytechnique F\'ed\'erale de Lausanne, 1015 Lausanne, Switzerland}
\thanks{These two authors contributed equally}

\author{Michele Ceriotti}
\email{michele.ceriotti@epfl.ch}
\affiliation{Laboratory of Computational Science and Modeling, Institut des Mat\'eriaux, \'Ecole Polytechnique F\'ed\'erale de Lausanne, 1015 Lausanne, Switzerland}

\newcommand{\mc}[1]{{\color{blue}#1}}
\newcommand{\fb}[1]{{\color{teal}#1}}

\date{\today}%

\begin{abstract}
Machine-learning models based on a point-cloud representation of a physical object are ubiquitous in scientific applications and particularly well-suited to the atomic-scale description of molecules and materials. 
Among the many different approaches that have been pursued, the description of local atomic environments in terms of their neighbor densities has been used widely and very succesfully. 
We propose a novel density-based method which involves computing ``Wigner kernels''. These are fully equivariant and body-ordered kernels that can be computed iteratively with a cost that is independent of the radial-chemical basis and grows only linearly with the maximum body-order considered. This is in marked contrast to feature-space models, which comprise an exponentially-growing number of terms with increasing order of correlations.
We present several examples of the accuracy of models based on Wigner kernels in chemical applications, for both scalar and tensorial targets, reaching state-of-the-art accuracy on the popular QM9 benchmark dataset, and we discuss the broader relevance of these ideas to equivariant geometric machine-learning.

\end{abstract}

\maketitle

Machine-learning techniques are widely used to perform tasks on 3D objects, from pattern recognition and classification to property prediction.\cite{gumh+01imr,guo+21tpami,wu+19ccvpr,li+21tnnls} 
In particular, different flavors of geometric machine learning\cite{bron+21arxiv} have been used widely in applications to chemistry, biochemistry and condensed-matter physics.\cite{gain+20nm,carl+19rmp,ceri+21cr} Given the coordinates and types of atoms seen as a point cloud, ML models act as a surrogate for accurate electronic-structure simulations, predicting all types of atomic-scale properties that can be obtained from quantum mechanical calculations.\cite{behl-parr07prl,bart+10prl,rupp+12prl,gilm+17icml,broc+17nc,ceri22mrsb} These include scalars such as the potential energy, but also vectors and tensors, that require models that are covariant to rigid rotations of the system.\cite{bere+15jctc,glie+17prb} 

In this context, \emph{body-ordered} models have emerged as an elegant and accurate way of describing how the behavior of a molecule or a crystal arises from a hierarchy of interactions between pairs of atoms, triplets, and so on -- a perspective that has also been widely adopted in the construction of traditional physics-based interatomic potentials.\cite{finn-sinc84pma,hors+96prb2,medd+15jcp,sanc+84pa}
By only modeling physical interactions up to a certain body order, these methods generally achieve low computational costs. Futhermore, since low-body-order interactions are usually dominant, focusing machine-learning models on their description also leads to excellent accuracy and data-efficiency. Several body-ordered models have been proposed for atomistic machine learning. While most work has been focused on relatively simple linear models\cite{drau19prb, duss+22jcp, niga+20jcp}, neural-network-like body-ordered models have also been explored\cite{bata+22arxiv}, and several classes of equivariant neural networks\cite{thom+18arxiv,ande+19nips,bazt+22ncomm} can be interpreted in terms of the systematic construction of hidden features that are capable of describing body-ordered symmetric functions\cite{niga+22jcp2,bata+22arxiv}.

Kernel methods have also been very popular in the field of atomistic chemical modeling\cite{bart+10prl,rupp+12prl,chmi+17sa,fabe+18jcp,gris+18prl,glie+18prb}, as they provide a good balance between the simplicity of linear methods and the flexibility of non-linear models. 
In most cases\cite{deri+21cr} they are used in an invariant setting, and the kernels -- although built from body-ordered components -- are subject to manipulations that incorporate higher-order terms in a non-systematic way.  
In principle, one could build equivariant, body-ordered kernels in terms of scalar products of the corresponding body-ordered descriptors. However, doing so would be impractical, as one would have to pay the price of evaluating a large number of features to then combine them into linear kernels that offer no additional descriptive power than the features themselves. 

In this work, we discuss an alternative approach to build body-ordered equivariant kernels in an iterative fashion. The iterations are performed in kernel space, and they therefore entirely avoid the definition of a basis to expand the radial and compositional (chemical element) descriptors of each atomic environment, along with the associated scaling issues. We demonstrate the excellent accuracy that is exhibited by these ``Wigner kernels" in the prediction of scalar and tensorial properties of molecular systems, including the cohesive energy of transition metal clusters and high-energy molecular configurations, and both energetics and molecular dipole moments for the QM9 dataset. 

\section{Related work}\label{sec:related-work}

The majority of machine-learning models for the prediction of atomistic properties rely on the representation of atom-centered environments. 
This ensures invariance of the predictions with respect to translations, and it results in a decomposition of the target properties into atomic contributions. Such atom-centered representations are often computed starting from the definition of local atomic densities around the atom of interest, which makes the predictions invariant with respect to permutation of atoms of the same chemical element. 
The locality of the atomic densities is often enforced via a finite cutoff radius within which they are defined, and it results in models whose cost scales linearly with the size of the system. 
Using discretized atomic densities has been linked to much increased computational efficiency in the evaluation of high-order descriptors, because they allow to compute them while avoiding sums over clusters of increasing order. This is sometimes referred to as the \emph{density trick}.\cite{vand+20mlst,musi+21cr}

\paragraph{Smooth overlap of atomic positions.}
Perhaps the oldest model to use the density trick is kernel-based SOAP-GPR\cite{bart+13prb}, which evaluates a class of 3-body invariant descriptors and builds kernels as their scalar products. Higher-body-order invariant interactions are generally included, although not in a systematic way, by taking integer powers of the linear kernels. SOAP-GPR has been used in a wide variety of applications.\cite{deri+21cr}

\paragraph{Symmetry-adapted GPR.}
 SA-GPR is an equivariant generalization of SOAP-GPR which aims to build equivariant kernels from ``$\lambda$-SOAP'' features.\cite{gris+18prl} In practice, these kernels are built as products of a linear low-body-order equivariant part and a non-linear invariant kernel that incorporates higher-order correlations. Hence, as in the SOAP-GPR case, the resulting kernels are not strictly body-ordered, and they offer no guarantees of behaving as universal equivariant approximators.

\paragraph{N-body kernel potentials}
In constrast, Ref.~\citenum{glie+18prb} introduces density-based body-ordered kernels, and it proposes their analytical evaluation for low body orders. Nonetheless, these kernels are exclusively invariant, and the paper proposes a strategy based on approximate symmetrization as the only viable strategy to compute kernels of arbitrarily high body-order.

\paragraph{Atomic cluster expansion/Moment tensor potentials/N-body iterative contraction of equivariants}
The atomic cluster expansion (ACE)\cite{drau19prb} and the related variants MTP\cite{shap16mms} and NICE\cite{niga+22jcp} consist of linear models based on a systematic hierarchy of equivariant body-ordered descriptors. These are obtained as discretized and symmetrized atomic density correlations, which are themselves simply tensor products of the atomic densities. Although several contraction and truncation schemes have been proposed,\cite{duss+22jcp, trace} in principle the ACE feature space grows exponentially with the maximum body order of the expansion.

\paragraph{Equivariant neural networks}
Finally, equivariant neural networks\cite{thom+18arxiv,ande+19nips,bazt+22ncomm,bata+22nips} have become ubiquitous in recent years, and they represent the state of the art on many atomic-scale datasets. Most, but not all\cite{allegro}, incorporate message-passing schemes.  
Equivariant architectures can be seen as a way to efficiently contract the exponentially large feature space of high body order density correlations.\cite{niga+22jcp2} Even though the target-specific optimization of the contraction weights gives these models great flexibility, they still rely on an initial featurization based on the expansion of the neighbor density on a local basis, and can only span a heavily contracted portion of the high-order correlation basis.

\section{Methods}

\subsection{(Symmetry-adapted) Kernel ridge regression}\label{sec:krr}

Throughout this work, we will employ Kernel ridge regression (KRR) to fit atomistic properties. In this context, kernels are defined between any two atomic-scale structures, so that the kernel $\krn(A,A')$ represents a similarity measure between structures $A$ and $A'$.
%
%
As mentioned in Sec. \ref{sec:related-work}, it is common practice -- rooted in physical approximations\cite{prod+05pnas} and usually beneficial to the transferability of the model -- to use atom-centered decompositions of the physical properties of a structure. In that case, the structure-wise kernels are decomposed into atom-pair contributions:\cite{de+16pccp}
\begin{equation}
   \krn(A, A') = \sum_{i \in A} \sum_{i' \in A'} \krn(A_i, A'_{i'}),
\end{equation}
where $i$ runs over all atoms in structure $A$, $i'$ runs over all atoms in structure $A'$, and $A_i$, $A'_{i'}$ denote the atomic environments around atoms $i$ and $i'$ (usually spherical neighborhoods of the central atoms with radius $\rcut$). 

As shown in Refs.~\citenum{glie+17prb} and \citenum{gris+18prl}, KRR can be extended to the prediction of atomistic properties that are equivariant with respect to symmetry operations in $SO(3)$ (3D-rotations $\hat{R}$). In order to build a symmetry-adapted model that is suitable for a property $y^\mu_\lambda$ that transforms like a set of spherical harmonics with angular momentum degree $\lambda$, it is necessary to employ tensorial kernels $\krn^\lambda_{\mu\mu'}$, and a symmetry-adapted regression ansatz
\begin{equation}\label{eq:sa-gpr}
\tilde{y}^\mu_\lambda(B) = 
\sum_A \sum_{\mu'} \krn^\lambda_{\mu\mu'}(B, A) \, c^{\mu'}_A.
\end{equation}
The $\krn^\lambda_{\mu\mu'}$ kernels must obey the property
\begin{equation}\label{eq:sa-krn-condition}
\krn^\lambda_{\mu \mu'}(\hat{R} A_i, \hat{R}' A'_{i'}) = 
\sum_{m m'} D^\lambda_{\mu m}(\hat{R}) D^\lambda_{\mu' m'}(\hat{R}') 
\krn^\lambda_{m m'}(A_i, A'_{i'}),
\end{equation}
where $D^{\lambda}_{\mu \mu'}(\hat{R})$ is the Wigner D-matrix associated with the rotation $\hat{R}$.
In practice, most established invariant models use some type of low-rank approximation of the kernel matrix, which results in a more favorable scaling with system size in training and predictions. See Ref.~\cite{deri+21cr} for a recent review on kernel methods applied to atomistic problems. 

\subsection{Atomic densities and body-ordered kernels}\label{sec:densities-kernels}

A broad class of atomistic ML frameworks can be formulated in terms of discretized correlations of an atomic neighbor density defined within each environment\cite{will+19jcp,drau19prb,musi+21cr,niga+22jcp2}, which can be written as
\begin{equation}\label{eq:density-expansion}
    \rho_i (\bx) = \sum_{j \in A_i} g (\bx - \br_{ji}) \, f_\mathrm{cut}(r_{ji}).
\end{equation}
Here, $j$ runs over all neighbors in $A_i$, $g$ is a three-dimensional Gaussian function, and $f_{\mathrm{cut}}$ is a cutoff function which satisfies $f_{\mathrm{cut}}(\rcut) = 0$, so that the $A_i$ neighborhoods are effectively restricted by a cutoff radius $\rcut$. This is necessary in order to obtain predictions that are simultaneously computationally cheap and continuous.\cite{musi+21jcp}

These densities can be used to define kernels that fulfill the equivariance condition~\eqref{eq:sa-krn-condition} 
\begin{equation}\label{eq:body-ordered-kernels}
    \krn^{\nu,\lambda}_{\mu\mu'}(A_i, A'_{i'}) = \int \D{\hat{R}}
\, D^{\lambda}_{\mu \mu'}(\hat{R})
\left( \int\rho_i (\bx) \, \rho_{i'}(\mathbf{R}^{-1} \bx) \, \D{\bx} \right)^{\nu},
\end{equation}
where $\nu$ will be referred to as the correlation order of the kernel, and the other symbols carry the same meaning as in Eq. \ref{eq:sa-krn-condition}. 
Similar kernels, with $\nu = 2$, have been used to machine learn tensorial properties of atomistic systems in Ref.~\citenum{gris+18prl}. The kernels in Eq. \ref{eq:body-ordered-kernels} contain correlated information about at most $\nu$ neighbors in each atomic neighborhood ($A_i$ and $A'_{i'}$). 
This is because the density expansion in Eq. \ref{eq:density-expansion} is a simple sum over neighbors, and it is raised to the power of $\nu$, while all other operations (the inner integral and the rotation) are linear. As a result, these kernels are intrinsically body-ordered: $\krn^{\nu,\lambda}_{\mu\mu'}$ can describe physical interactions up to body order $\nu+1$ (the center of the representation and $\nu$ neighbors), but not higher.

\subsection{Wigner kernels through Wigner iterations}

As detailed in Appendix \ref{app:features}, symmetry-adapted kernels of the form given in~\eqref{eq:body-ordered-kernels} can be computed by first evaluating body-ordered equivariant \emph{representations} (in the form of discretized correlations of the neighbor density~\eqref{eq:density-expansion}) and then computing their scalar products. 
However, doing so is impractical for $\nu>2$: on one hand, kernel regression is then equivalent to linear regression on the starting features, so that the calculation of the kernel is an unnecessary step; on the other, the number of features one needs to compute to evaluate the kernel without approximations grows exponentially with $\nu$.

Our main result, which we will refer to as a Wigner iteration (and derive in Appendix \ref{app:derivation}), is that high-$\nu$ kernels can be computed following an alternative route, by combining lower-order kernels iteratively:
\begin{multline}\label{eq:wigner-iteration}
    \!\!\!\krn_{\mu \mu'}^{(\nu+1),\lambda} (A_i, A'_{i'})  = 
    \!\!\!  \sum_{\substack{l_1 m_1 m_1' \\ l_2 m_2 m_2'}} \!\!\!
\cg{l_1m_1}{l_2m_2}{\lambda\mu}\krn_{m_1 m_1'}^{\nu, l_1} (A_i, A'_{i'})  \\
\times \krn_{m_2 m_2'}^{1, l_2} (A_i, A'_{i'}) 
 \cg{l_1m_1'}{l_2m_2'}{\lambda\mu'},
\end{multline}
where $\cg{l_1m_1}{l_2m_2}{\lambda\mu}$ are Clebsch-Gordan coefficients.

In order to initialize the iterations in Eq. \ref{eq:wigner-iteration}, only the $\nu = 1$ equivariant kernels $\krn_{\mu \mu'}^{1, \lambda}$ are needed. These are relatively simple to define as a double sum over neighbors using Eqs. \ref{eq:density-expansion} and \ref{eq:body-ordered-kernels}:
\begin{multline}\label{eq:nu1-kernels} 
\!\!\!\krn_{\mu \mu'}^{1, \lambda} (A_i, A'_{i'}) = \delta_{a_i a_{i'}} 
    \int \D{\hat{R}} \, D^{\lambda}_{\mu \mu'}(\hat{R})
  \sum_{\substack{j \in A_i\\ j' \in A'_{i'}}} \delta_{a_j a_{j'}} \\[-0.5em]
\!\!\!\times f_\mathrm{cut}(r_{ji}) \, f_\mathrm{cut}(r_{j'i'})
\int \!  g(\bx - \br_{ji}) \, g (\bx - \mathbf{R}^{-1}\br_{j'i'}) \,  \D{\bx},
\end{multline}
where the $\delta_{a_ia_{i'}}$ term simply indicates that kernels between atoms of different chemical species are set to zero.  

The integrals in Eq. \ref{eq:nu1-kernels} could be evaluated analytically, although in practice, and for simplicity, we compute them numerically as scalar products of an atom-centered density expansion, as explained in Appendix \ref{app:features}. 
Finally, equivariance with respect to inversion is discussed in Appendix \ref{app:inversion}, and it results in the incorporation of a parity index $\sigma$, so that the full notation for an $O(3)$-equivariant kernel is $\krn_{\mu \mu'}^{\nu, \lambda\sigma}$.
Although not necessary to initialize the iterations, we also define one-body, $\nu = 0$ kernels as 
\begin{equation}
    \krn_{\mu \mu'}^{0, \lambda \sigma}(A_i, A'_{i'}) = \delta_{\lambda 0} \delta_{\sigma 1} \delta_{a_i a_i'},
\end{equation}
that describe similarity of two environments based exclusively on the nature of the central atom, and will be useful in Section~\ref{app:implementation} to define non-linear kernel functions.

\subsection{Scaling and computational cost}\label{sec:scaling}

The evaluation of high-$\nu$ Wigner kernels as scalar products of equivariant features (see Appendix B) would require aggressive truncation as a consequence of the exponential scaling of the equivariant feature set size as a function of $\nu$.\cite{niga+20jcp,duss+22jcp}. 
The main advantage of the Wigner iteration~\eqref{eq:wigner-iteration} is that the body-ordered ``Wigner kernels'' in Eq.~\ref{eq:body-ordered-kernels} can be calculated in a fully-converged radial-element space at no additional cost. 
The scaling with respect to the maximum neighbor correlation order $\nu_\text{max}$, the number of elements $a_\text{max}$, radial basis cutoff $n_\text{max}$, maximum angular momentum $\lambda_\text{max}$ and train set size $n_\text{train}$ is discussed in Appendix~\ref{app:scaling}, where we show that Wigner kernels model have a cost that is independent on $a_\text{max}$ and $n_\text{max}$, linear in $\nu_\text{max}$, polynomial in $\lambda_\text{max}^7$. For a naive implementation, inference is linear in $n_\text{train}$, although it would be comparatively simple to avoid this scaling implementing a sparse KRR framework.

Traditionally, linear body-ordered methods such as ACE exhibit exponential scaling of the feature space with increasing $\nu$, which can be expressed approximately as $(a_\mathrm{max} \, n_\mathrm{max}\, \lambda_\text{max} )^\nu$. Even if this scaling can be (and usually is) mitigated with several heuristic or data-driven approximations\cite{will+19jcp, niga+20jcp, trace}, the possibility of computing the value of the kernel without any truncation, and eliminating completely the exponential scaling with $\nu_{\text{max}}$ is particularly appealing.
However, the steep scaling with $\lambda_{\mathrm{max}}$ is a potential drawback of Eq.~\eqref{eq:wigner-iteration}: while Wigner iterations scale as $\lambda_{\mathrm{max}}^7$, traditional SO(3)-symmetrized products scale as $\lambda_{\mathrm{max}}^5$ (see e.g. the five angular indices of Eq. 46 in Ref. \cite{musi+21cr}). 
Additionally, most ACE implementations reduce the number of radial basis functions for high $l$ or $\lambda$ values, which is physically motivated by smoothness arguments, and which reduces the computational cost as fewer features enter the more expensive high-$\lambda$ components of SO(3)-symmetrized products. 
The computation of Wigner kernels in Eq. \ref{eq:wigner-iteration} instead forces the same $n_\text{train}$-sized basis on all $\lambda$ channels, which aggravates the impact of the steep scaling. Fortunately, as we shall see, excellent performance can be achieved with low $\lambda_\text{max}$.

\section{Results}\label{sec:results}

Having discussed the formulation and the theoretical scaling of Wigner kernels, we now proceed to assess their behavior in practical regression tasks, focusing on applications to atomistic machine learning.
We refer the reader to Appendix~\ref{app:implementation} for a discussion of the implementation details, and to Appendix~\ref{app:hypers} for a list of the hyperparameters of the models. 
We consider three cases that allow us to showcase the accuracy of our framework: a system that is expected to exhibit strong many-body effects, one that requires high-resolution descriptors, a problem that involves a tensorial target, and finally  a classical benchmark dataset for organic molecules.
We use learning curves as a tool to assess the balance between parameterization complexity and generalization power of different models\cite{huan-vonl16jcp,bart+17sa,corna+23mlst}. 
Learning curves that show saturation of the validation error with train set size indicate that the kernel does not have sufficient information to incorporate new data on the structure-property relations, whereas an algebraic decay of the error with train-set size indicates that the model is data-limited. 

\subsection{Gold cluster dataset}

The scaling properties of the Wigner kernel model discussed in Sec.~\ref{sec:scaling} make it especially advantageous for systems requiring a high-body-order description of the potential energy surface. Metallic clusters often exhibit non-trivial finite-size effects due to the interplay between surface and bulk states~\cite{li+13jpcl}, and have been used in the past as prototypical benchmarks for many-body ML models\cite{zeni+18jcp}.
As a particularly challenging test case, we built a dataset of $105\,092$ uncorrelated structures of gold clusters of different size, extracting them from the long molecular dynamics trajectories performed in Ref.~\citenum{gold+19prm}. The complete MD trajectories are available in an online repository~\cite{nomad16a}. 

\begin{figure}[tbp]
    \centering
    \includegraphics[width=\linewidth]{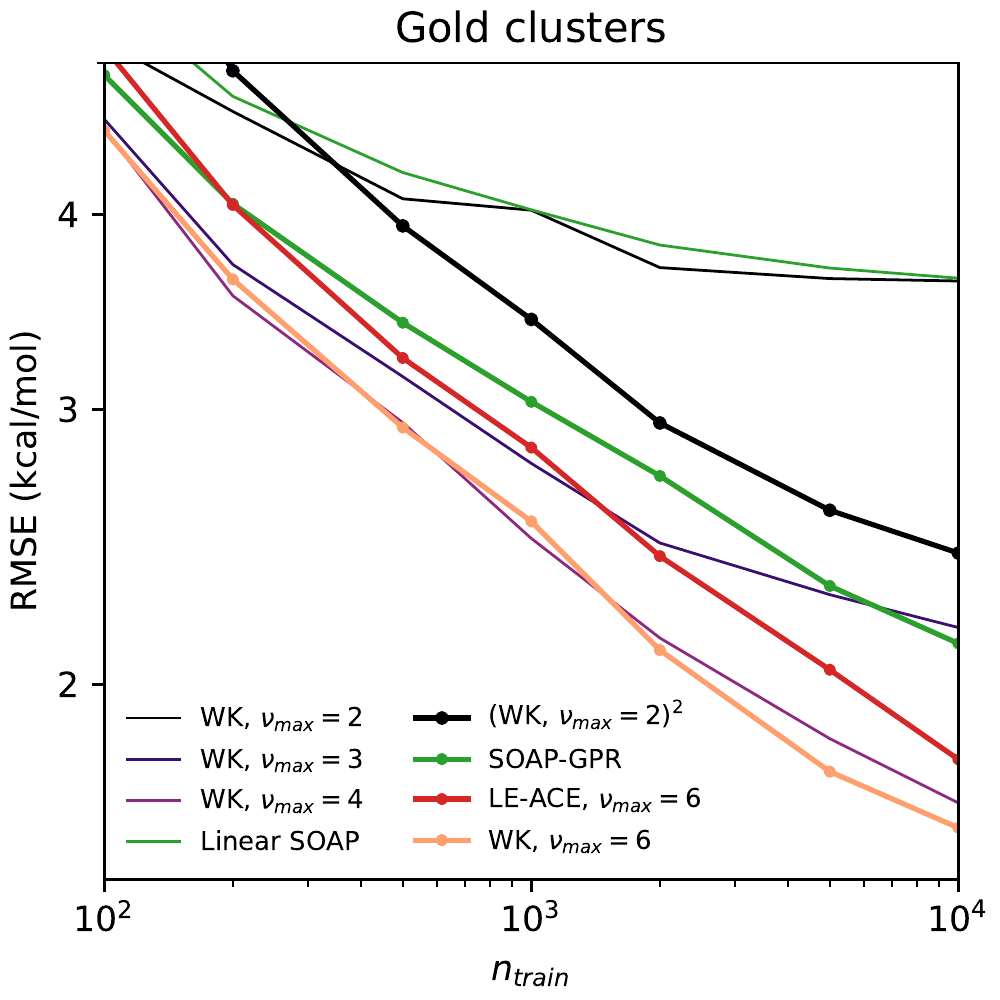}
    \caption{Learning curves for the electronic free energy of gold clusters. Different curves correspond to invariant Wigner kernels of increasing body order, as well as a construction where 
    a linear combination of Wigner kernels up to $\nu_\text{max} = 2$ is squared. A linear SOAP\cite{bart+13prb} model, a SOAP-GPR\cite{deri+21cr} model built with a squared kernel, and a LE-ACE\cite{bigi+22jcp} model are also shown. The hyperparameters for all models are discussed in Appendix \ref{app:hypers}.
    }
    \label{fig:gold}
\end{figure}

The need for high-body-order terms is clear when comparing results for models based on exponential WKs truncated at different orders of $\nu$ (Fig.~\ref{fig:gold}). $\nu=2$ and (to a lesser extent) $\nu=3$ models result in saturating learning curves, although for the maximum dataset size we consider $\nu=4$ kernels are almost as accurate as the highest case we consider, with $\nu=6$. 
A comparison with SOAP-based models reveals the likely source of the increased performance of the Wigner kernels. Indeed linear SOAP, which is a $\nu_{\max}=2$ model, shows very similar performance to its WK counterpart. The same is true for squared-kernel SOAP-GPR, which closely resembles the learning curve of a Wigner kernel construction for which $\nu_{\max}=2$ and the resulting kernels are squared - the difference probably due to the different functional form of the two kernels, and the presence of higher-$l$ components in the density for SOAP-GPR. 
A true $\nu_\text{max} = 4$ kernel, that incorporates \emph{all} five-body correlations, significantly outperforms both squared-kernel learning curves, demonstrating the advantages of explicit body-ordering.
We conclude with a comparison between the $\nu_\text{max} = 6$ WKs and $\nu_\text{max} = 6$ Laplacian-Eigenbasis (LE) ACE models. For the latter, we used the same radial transform presented in Ref. \cite{bigi+22jcp}, and we optimized its single hyperparameter. Although it might be possible to further tune the performance of LE-ACE by changing the functional form of the radial transform altogether, the comparison with the Wigner kernel learning curve suggests that the kernel-space basis employed in the Wigner kernels might be advantageous in geometrically inhomogeneous datasets such as this one.

\begin{figure}[h]
    \centering
    \includegraphics[width=\linewidth]{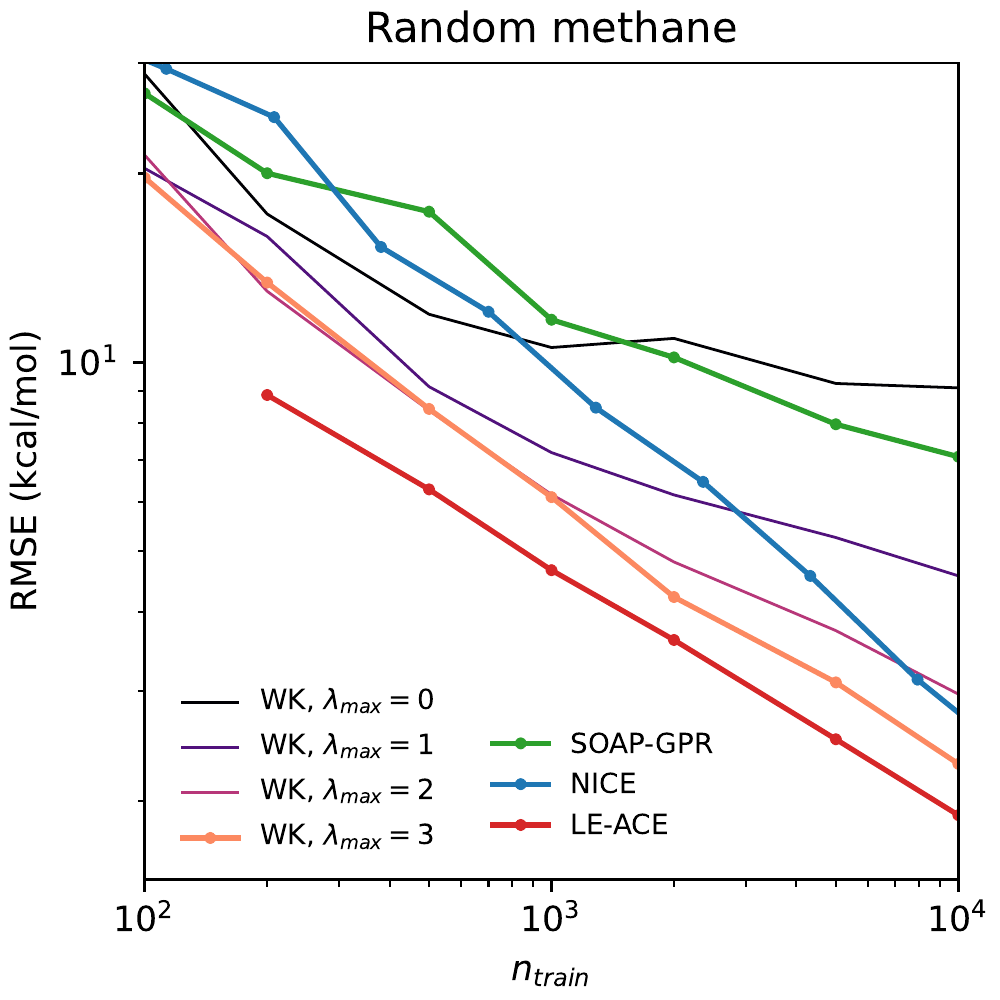}
    \caption{Learning curves for the energy of random \ce{CH4} configurations, comparing different models. The LE-ACE and NICE curves are from Refs. \citenum{bigi+22jcp} and \citenum{niga+20jcp}, respectively. Hyperparameters for all other models are discussed in Appendix \ref{app:hypers}.   
    We note that the REANN neural network\cite{zhan+22jcp} achieved higher accuracy on this dataset by also learning from forces. While, at present, our implementation does not allow to train with target gradients, preliminary results show similar accuracy between LE-ACE and REANN when both are trained on energies and forces. 
    }
    \label{fig:methane}
\end{figure}

\subsection{Random methane dataset}\label{sec:methane}

As a second example, we test Wigner kernels on a dataset of random gas-phase \ce{CH4} structures,\cite{pozd+20prl,matcloud20a} which we expect to be very challenging for the proposed model. Firstly, this dataset is intrinsically limited in body-order: a $\nu_\text{max}=4$ model is in principle sufficient to describe all interatomic correlations in a \ce{CH4} configuration. Secondly, since the atomic positions are almost random, it is also less advantageous to use a kernel basis, that is naturally adapted to the structures that are part of the training set, in comparison with a linear expansion in terms of explicit real-space density correlations. 
It is also important to consider that this dataset requires very careful convergence of the angular basis,\cite{niga+22jcp2,bigi+22jcp} which is problematic in view of the steep $\lambda_\text{max}$ scaling of Wigner iterations. 
The difficulty in approximating the potential energy surface is apparent in how both a SOAP-GPR model (with $l_\text{max} = 6$) and a model based on an adaptive contraction of high-order correlations using the NICE framework\cite{niga+20jcp} (with $\lambda_\text{max} = 10$) perform noticeably worse than one that uses a highly optimized Laplacian eigenvalue basis, that includes terms up to $l=20$. 

With all these potential problems, Wigner kernels achieve a remarkable level of accuracy, outperforming SOAP-GPR and NICE, and being competitive with LE-ACE despite using only $\lambda_\text{max}=3$. Similar effects have been noticed in many recent efforts to machine-learn interatomic potentials \cite{nequip, bata+22arxiv, allegro, mace}. 
By providing a functional form that spans the full space of density correlations at a given level of angular truncation, Wigner kernels can help rationalize why low-$\lambda_\text{max}$ models can perform well. 
For starters, one could hypothesize that a high angular resolution is not necessary to approximate local energies, because the most important energetic contributions come from the close field, where a low-$l$ fit can still provide good resolution due to the smaller 3D volume element. This interpretation is consistent with previous observations on this same data set,\cite{niga+22jcp2} showing that the need for very high angular components is particularly strong when using only the C atom as center, while a multi-center ansatz such as the one we use here makes it possible to use low-order H-centered correlations to describe the \ce{H-H} interactions occurring in dissociated configurations.
More importantly, due to the form of the Wigner iterations, $\krn^{(\nu)} $ does not report \emph{exclusively} on $(\nu+1)$-body correlations, but also on all lower-order ones, and the tensor-product form of the kernel space incorporates higher frequency components in their functional form, much like $\sin^2 \omega x$ contains components with frequency $2\omega$. 
We demonstrate this phenomenon in Appendix~\ref{app:angular-scans} by decomposing the angular dependence of high-$\nu$ kernels into their frequency components. The combined effect of increasing $\nu$ complicates the interpretation of ablation studies, making it difficult to disentangle the effects of correlation order and of angular resolution.
However, in practice it facilitates the convergence of regression models, and explains how aggessively-truncated equivariant ML models\cite{schu+21icml,bazt+22ncomm} can achieve high accuracy in the prediction of interatomic potentials. 

\subsection{QM9 dataset}

Wigner kernels also avoid the unfavorable scaling of traditional body-ordered models with respect to the number of chemical elements in the system. This property is particularly useful when dealing with chemically diverse datasets. An example is that of the popular QM9 dataset, which contains 5 elements (H, C, N, O, F). 

We build KRR models for two atomic-scale properties within this dataset, and, to illustrate the transferability of our model, we use the same hyperparameters for both fits (see Appendix \ref{app:fit-details}).

\begin{figure}[tbp]
    \centering
    \includegraphics[width=\linewidth]{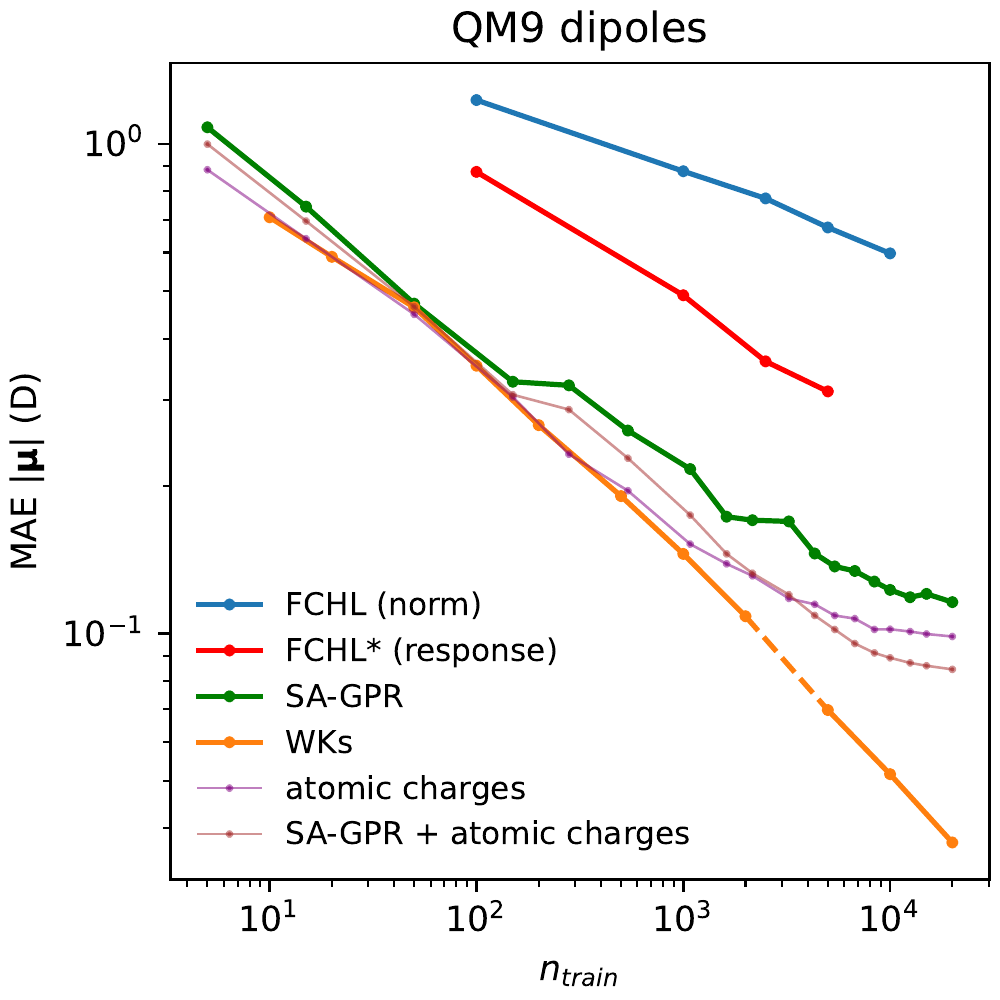}
    \caption{Learning curves for the prediction of molecular dipole moments in the QM9 datasets. Different curves correspond to FCHL kernels\cite{fabe+18jcp}, the dipole models presented in Ref.~\citenum{veit+20jcp}, and Wigner kernels. It should be noted how the models that use atomic charges can account for the macroscopic component of the dipole moment that arises due to charge separation, while the others predict dipole moments as a sum of local atom-centered contributions. 
    The dashed line in the WK learning curve represents a change in the fitting procedure: the points before the dashed line are obtained as highlighted in Sec. \ref{sec:exp-kernels}, while the points after the dashed line are obtained with the same cross-validation procedure, but using a less expensive 2-dimensional grid search instead of dual annealing (note that $c_0$ does not need to be considered in a covariant fit as $\krn_{\mu \mu'}^{0, \lambda \sigma} = 0$). The accuracy of the model does not seem to be affected by this change. }
    \label{fig:dipoles}
\end{figure}

\subsubsection{Molecular dipoles}
We begin the investigation with a covariant learning exercise. This consists of learning the dipole moment vectors $\boldsymbol{\mu}$ of the molecules in the QM9 dataset.\cite{veit+20jcp} 
In the small-data regime, Wigner kernels have a similar performance to that obtained by optimized $\lambda$-SOAP kernels in Ref.~\citenum{veit+20jcp}, but avoid completely the saturation for larger train set size (Fig.~\ref{fig:dipoles}). 
Even though here we perform exclusively predictions on molecules of consistent size -- whereas the physically-motivated charge-separation model used in Ref.~\citenum{veit+20jcp} is expected to be most advantageous when extrapolating to larger molecules -- the improved performance of a model that is based exclusively on local dipole contributions is a clear indication of the higher descriptive power that is afforded by the use of a full body-ordered equivariant kernel, as opposed to the combination of linear covariant $\nu=2$ kernels and non-linear scalar kernel that is used in current applications of SA-GPR. 

\subsubsection{Energies}

\begin{figure}[tbp]
    \centering
    \includegraphics[width=\linewidth]{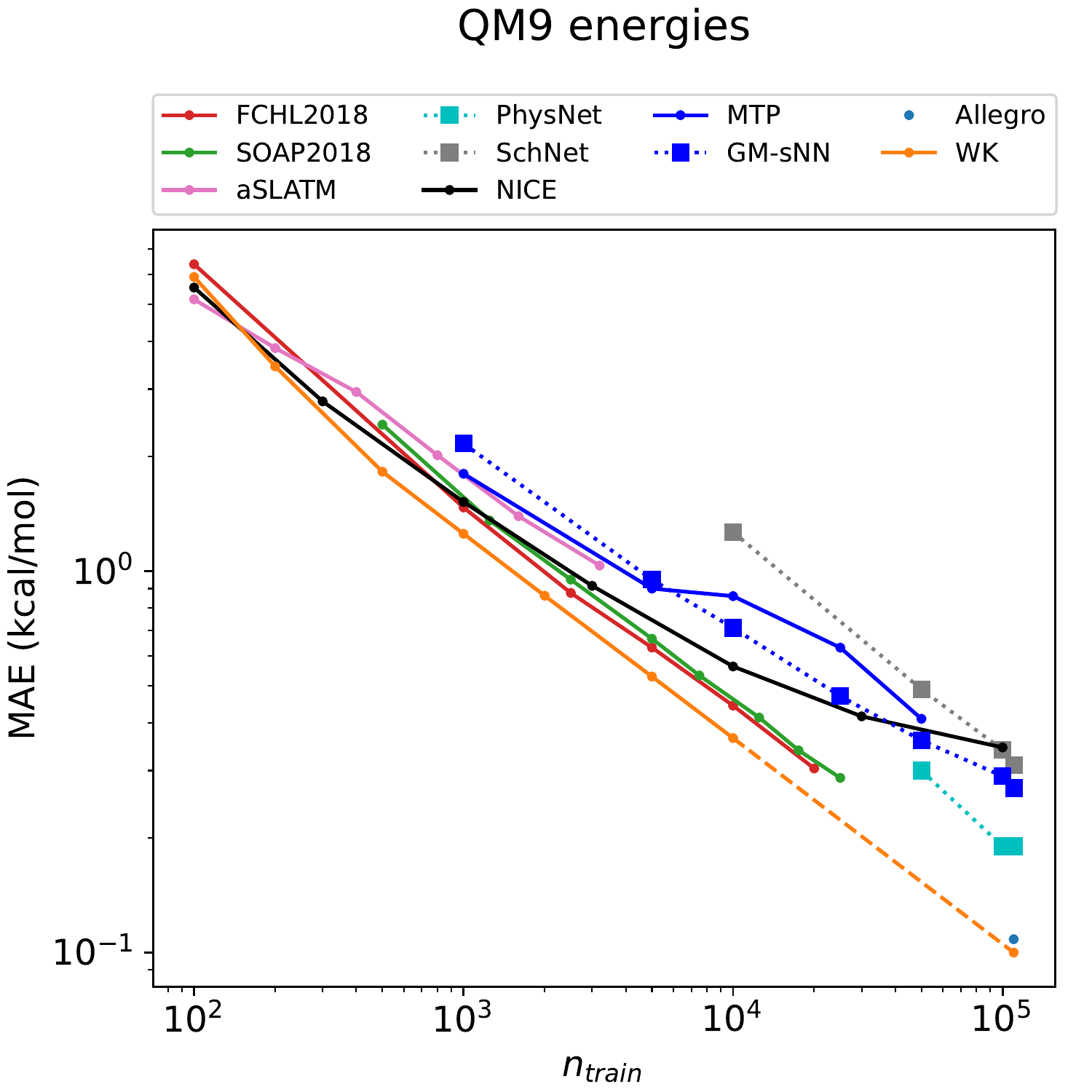}
    \caption{Selection of the best QM9 literature models for which learning curves are available: FCHL\cite{fabe+18jcp}, SOAP-GPR\cite{will+18pccp}, aSLATM\cite{huan-vonl16jcp}, PhysNet\cite{unke-meuw19jctc}, SchNet\cite{schu+18jcp}, NICE\cite{niga+20jcp}, MTP\cite{shap16mms}, GM-sNN\cite{zave-kast20jctc}. Allegro\cite{allegro} is also included for completeness, as it represents the most accurate model on the \emph{full} QM9 dataset previous to this work. More literature models trained on the full dataset are shown in Table \ref{tab:full-qm9}. The dashed line in the WK learning curve represents a change in the fitting procedure. The points to its left are obtained by averaging 10 runs with random train/test splits, and cross-validation is conducted within the training set as described in Sec. \ref{sec:exp-kernels}. Instead, for consistency with the literature models trained on the full QM9 dataset (Table \ref{tab:full-qm9}), the last point is averaged over 16 random train/validation/test splits where validation is conducted on a dedicated validation set via a grid search.}
    \label{fig:qm9}
\end{figure}

Finally, we test the Wigner kernel model on the ground-state energies of the QM9 dataset. The corresponding learning curves are shown in Fig. \ref{fig:qm9}.
Wigner kernels significantly improve on other kernel methods such as SOAP and FCHL in the low-data regime. As in the methane case, the WK model is truncated at a low angular frequency threshold ($\lambda_\text{max} = 3$). However, the corresponding learning curve shows no signs of saturation, possibly for the same reasons we highlighted in Sec. \ref{sec:methane}. Similarly, a relatively low maximum body order ($\nu_\text{max} = 4$) does not seem to impact the accuracy of the model, possibly because stable organic molecules have, with few exceptions, atoms with only up to four nearest neighbors.

On the full QM9 dataset, Wigner kernels also achieve state-of-the-art accuracy, as shown in the last point of the WK learning curve and in Table \ref{tab:full-qm9}. It is remarkable that it is still possible to achieve a 10\% improvement in regression performance for a dataset that has been used for benchmarks for a decade, and that kernel regression can outperform ``deep'' equivariant models. 
It is also worth mentioning that, similar to Allegro\cite{allegro}, our model is entirely local, and does not incorporate message-passing operation, which would simplify the parallelization of inference for large-scale calculations. 
The impressive performance of the newly proposed model on this exercise can be partly attributed to the nature of the QM9 dataset itself, which only contains molecules in their ground-state configurations. In this scenario, the definition of a potential energy surface as a function of real-space coordinates carries little meaning, and, formally, such a function is what most equivariant neural networks aim to approximate.
Instead, we hypothesize that kernel models based on a physically motivated similarity measure between structures are more suited to this kind of task, which, in real-world applications, would correspond to applications such as screening of pharmaceutical targets or prediction of chemical shifts from single equilibrium configurations. This stands in contrast to the other datasets we have investigated, which are better suited to assess the quality of a model in approximating a property surface for atomistic simulations.

\begin{table}[tbp]
    \centering
    \begin{tabular}{cc}
         \hline\hline
         Model & Test MAE (meV) \\
         \hline
         DimeNet++\cite{klicpera2020fast} & 6.3 \\
         SphereNet\cite{coors2018spherenet} & 6.3 \\
         ET\cite{tholke2022equivariant} & 6.2 \\
         NoisyNodes\cite{godwin2021simple} & 7.3 \\
         PaiNN\cite{schu+21icml} & 5.9 \\
         Allegro\cite{allegro} & 4.7 (0.2) \\
         Wigner Kernels & \textbf{4.3} (0.1) \\  
         \hline\hline
    \end{tabular}
    \caption{Performace comparison of the Wigner kernel model with a selection of the best literature models on the full QM9 dataset, as presented in Ref.~\citenum{allegro}. The Wigner kernel values are the mean and standard deviation of 16 runs on different random train/validation/test splits. In particular, the training set contains $110\,000$ random structures, the validation set another $10\,000$, and all the remaining QM9 structures constitute the test set, for consistency with Ref.~\citenum{allegro}.}
    \label{tab:full-qm9}
\end{table}

\section{Conclusions}

In this work, we have presented the Wigner iteration as a practical tool to construct rotationally equivariant ``Wigner kernels'' for use in symmetry-adapted Gaussian process regression on 3D point clouds. We have then applied them to machine learn the atomistic properties of molecules and clusters. 
The proposed kernels are explicitly body-ordered -- i.e. they provide explicit universal approximation capabilities for properties that depend simultaneously on the correlations between the positions of $\nu+1$ points -- and can be thought as the kernels corresponding to several families of body-ordered descriptors. 
Whereas the full feature-space evaluation of body-ordered models leads to an exponential increase of the cost with $\nu$, a kernel-space evaluation is naturally adapted to the training structures, and it avoids the explosion in the number of equivariant features that arises from the use of an explicit radial-chemical basis. 
The scaling properties of the Wigner iterations make the new model particularly suitable for datasets which are chemically diverse, which are expected to contain strong high-body-order effects, and/or which involve a very inhomogeneous distribution of molecular geometries so that a kernel-space model can provide a more efficient parameterization than a feature-space one.

The results of our benchmarks demonstrate that KRR models based on Wigner kernels show excellent performance for a variety of different atomistic problems. 
The energetics of gold clusters converge systematically with the body order of the expansion, and it appears to be essentially converged at $\nu_\text{max}=6$. This type of convergence is consistent with the expectation of substantial contributions from high-order terms in small metallic clusters, and the comparison with non-systematic kernels such as polynomial SOAP demonstrates the advantage in models that span the full basis of density correlations.
The results for a random methane dataset, which does not have the high-body-order behavior of gold clusters and is expected to require a highly-converged description of angular correlations, suggest that Wigner kernels incorporate high-resolution basis functions even when they are built with a moderate angular momentum threshold, which is reassuring given the steep scaling of the computational cost with $\lambda_\text{max}$.
Finally, the chemically diverse QM9 dataset allows us to showcase the state-of-the-art performance of the proposed model when performing vectorial learning of the molecular dipole moments and, perhaps even more remarkably, when predicting atomization energies. 
Achieving an accuracy of 0.1 kcal/mol for a dataset based on DFT energies is of little practical use, and we do not intend to overstate the importance of this result. However, the fact that a kernel model can still outperform extensively-tuned equivariant neural networks testifies to the importance of understanding the connection between body-ordered correlations, the choice and truncation of a feature-space basis, and the introduction of scalar non-linearities in kernel and equivariant models. 

Besides this fundamental role to test the complete-basis limit of linear density-correlation models, we believe it may be possible to also incorporate Wigner iterations into practical applications. 
The steep computational cost is largely due to the use of full KRR models - particularly given that in all our tests the use of an aggressive truncation of the angular iteration did not lead to a dramatic degradation in model performance. Using a sparse kernel formalism, possibly optimizing the active points, should make the model competitive with scalar Gaussian approximation models. 
Furthermore, the Wigner iteration could also be applied outside a pure kernel framework: from the calculation of non-linear equivariant functions, to the inclusion as a node in an equivariant architecture, the ideas we present here open up an original research direction in the construction of symmetry-adapted, physically-inspired models for chemistry, materials science, and more in general any application whose inputs can be conveniently described in terms of a 3D point cloud. 

\begin{acknowledgments}
The Authors would like to thank Jigyasa Nigam and Kevin Huguenin-Dumittan for stimulating discussions. %
MC and FB acknowledge support from the NCCR MARVEL, funded by the Swiss National Science Foundation (SNSF, grant number 182892). MC and SP acknowledge support from the Swiss Platform for Advanced Scientific Computing (PASC).
\end{acknowledgments}

\providecommand{\noopsort}[1]{}

\appendix
\clearpage
\section{Derivation of the Wigner iteration}\label{app:derivation}

In this Appendix, we derive Eq. \ref{eq:wigner-iteration}. In order to make the notation more compact, we define unsymmetrized versions of the equivariant kernels in Eq. \ref{eq:body-ordered-kernels} which only contain the inner integral:
\begin{equation}
    \krn^{\nu} (A_i, \hat{R}A_i') = \left( \int \rho_i (\bx) \, \rho_{i'}(\mbf{R}^{-1} \bx) \, \D{\bx} \right)^{\nu},
\end{equation}
from which it is straightforward to see that 
\begin{equation}\label{eq:nu_additivity}
    \krn^{\nu+\nu'} (A_i, \hat{R}A_i') = \krn^{\nu} (A_i, \hat{R}A_i') \, \krn^{\nu'} (A_i, \hat{R}A_i').
\end{equation}

According to its behavior upon the relative rotation of its two densities $\hat{R}$, the $\krn^{\nu}$ kernel can be decomposed into $\krn_{\mu \mu'}^{\nu, \lambda}$ contributions. Within the space of rotations $\hat{R}$, the latter kernels are effectively the expansion coefficients of $\krn^{\nu}$ in the basis of the Wigner D-matrices $D^{\lambda}_{\mu \mu'}(\hat{R})$, so that
\begin{equation}\label{eq:wigner_D_expansion}
    \krn^{\nu}(A_i, \hat{R} A_{i'}) = \sum_{\lambda \mu \mu'} \krn_{\mu \mu'}^{\nu, \lambda} (A_i, A_{i'}) \, D^{\lambda}_{\mu \mu'}(\hat{R})
\end{equation}
and
\begin{equation}\label{eq:wigner_D_reverse}
    \krn_{\mu \mu'}^{\nu, \lambda} (A_i, A_{i'}) = \int D^{\lambda}_{\mu \mu'}(\hat{R})^{*} \, \krn^{\nu}(A_i, \hat{R} A_{i'}) \, d\hat{R},
\end{equation}
which corresponds to Eq.~\eqref{eq:body-ordered-kernels} in the main text.

A simple combination of Eqs. \ref{eq:nu_additivity}, \ref{eq:wigner_D_expansion} and \ref{eq:wigner_D_reverse} leads to our main result: 
\begin{multline}\label{proof}
\!\!\!\krn_{\mu \mu'}^{\nu+1, \lambda} (A_i, A_{i'}) \stackrel{(\ref{eq:wigner_D_reverse})}{=} \!\int \!D^{\lambda}_{\mu \mu'}(\hat{R})^{*} \krn^{\nu+1}(A_i, \hat{R} A_{i'}) \, \D{\hat{R}} \stackrel{(\ref{eq:nu_additivity})}{=} \\ 
    \int D^{\lambda}_{\mu \mu'}(\hat{R})^{*} \, \krn^{\nu}(A_i, \hat{R} A_{i'}) \, \krn^{1}(A_i, \hat{R} A_{i'}) \, d\hat{R} \stackrel{(\ref{eq:wigner_D_expansion})}{=} \\
    \sum_{\substack{l_1 m_1 m_1'\\l_2 m_2 m_2'}} \krn_{m_1 m_1'}^{\nu, l_1} (A_i, A_{i'}) \, \krn_{m_2 m_2'}^{1, l_2} (A_i, A_{i'}) \\
    \int D^{\lambda}_{\mu \mu'}(\hat{R})^{*} \, D^{l_1}_{m_1 m_1'}(\hat{R}) \, D^{l_2}_{m_2 m_2'}(\hat{R}) \, d\hat{R} =  \\
    \sum_{\substack{l_1 m_1 m_1'\\l_2 m_2 m_2'}} C^{l_1l_2\lambda}_{m_1m_2\mu} \krn_{m_1 m_1'}^{\nu, l_1} (A_i, A_{i'}) \, \krn_{m_2 m_2'}^{1, l_2} (A_i, A_{i'}) \, C^{l_1l_2\lambda}_{m_1'm_2'\mu'}\,,
\end{multline}
where, in the last equality, we have used a well-known property of the Wigner D-matrices which relates them to the Clebsch-Gordan coefficients $C^{l_1l_2L}_{m_1m_2M}$.

\section{Density-correlation view of the Wigner iteration}\label{app:features}

A different derivation of the Wigner iteration (Eq. \ref{eq:wigner-iteration}) can reveal a direct connection with frameworks that operate in feature space (ACE\cite{drau19prb}, NICE\cite{niga+20jcp}). To achieve this, we will use the notation from Ref. \cite{niga+20jcp}. 
Starting from Eq. \ref{eq:body-ordered-kernels}, we obtain
\begin{multline}\label{scalar-product}
    \krn^{\nu, \lambda}_{\mu \mu'}(A_i, A_{i'}) = \int \D{\hat{R}}
    \, D^{\lambda}_{\mu \mu'}(\hat{R})^{*} \int\rho_i (\bx^\nu) \, \rho_{i'}((\mbf{R}^{-1} \bx)^\nu) \, \D{\bx^\nu} = \\
    \int \D{\hat{R}} \, D^{\lambda}_{\mu \mu'}(\hat{R})^{*} \, \sum_{LMq} \langle \rho_i^{\otimes \nu}; LM \vert q \rangle \, \hat{R} \langle q \vert \rho_{i'}^{\otimes \nu}; LM \rangle = \\
    \int \D{\hat{R}} \, D^{\lambda}_{\mu \mu'}(\hat{R})^{*} \, \sum_{LMq} \langle \rho_i^{\otimes \nu}; LM \vert q \rangle \, \sum_{M'} D^{L}_{MM'}(\hat{R}) \langle q \vert \rho_{i'}^{\otimes \nu}; LM' \rangle = \\
    \delta_{L\lambda} \delta_{M\mu} \delta_{M'\mu'} \sum_{LMq} \langle \rho_i^{\otimes \nu}; LM \vert q \rangle \, \sum_{M'} \langle q \vert \rho_{i'}^{\otimes \nu}; LM' \rangle = \\
    \sum_q \langle \rho_i^{\otimes \nu}; \lambda \mu \vert q \rangle \langle q \vert \rho_{i'}^{\otimes \nu}; \lambda \mu' \rangle.
\end{multline}
The first equality is a rearranged version of Eq. \eqref{eq:body-ordered-kernels}, the second is a change of basis from real space to a rotationally symmetrized $LMq$ basis, the third and fourth employ properties of the Wigner D-matrices, and the fifth follows immediately.
This equation shows that Wigner kernels correspond to scalar products between body-ordered density-correlation features, computed in the limit of a complete basis set expansion for the neighbor density. 

Now, using the iterative equivariant construction presented in Ref.~\citenum{niga+20jcp}, it can be noted that
\begin{multline}
    \krn^{\nu+1, \lambda}_{\mu\mu'}(A_i, A_{i'}) = \sum_q \langle \rho_i^{\otimes (\nu+1)}; \lambda \mu \vert q \rangle \langle q \vert \rho_{i'}^{\otimes (\nu+1)}; \lambda \mu' \rangle = \\
    \sum_{l_1 m_1 m_1'} \sum_{l_2 m_2 m_2'} \sum_{q'n} C^{l_1 l_2 \lambda}_{m_1 m_2 \mu} C^{l_1 l_2 \lambda}_{m_1' m_2' \mu'} \\
    \langle \rho_i^{\otimes \nu}; l_1 m_1 \vert q' \rangle \langle \rho_i^{\otimes 1}; l_2 m_2 \vert n \rangle
    \langle n \vert \rho_{i'}^{\otimes 1}; l_2 m_2' \rangle \langle q' \vert \rho_{i'}^{\otimes \nu}; l_1 m_1' \rangle = \\ 
    \sum_{\substack{l_1 m_1 m_1' \\ l_2 m_2 m_2'}} C^{l_1 l_2 \lambda}_{m_1 m_2 \mu} \krn^{\nu, l_1}_{m_1 m_1'}(A_i, A_{i'}) \krn^{1, l_2}_{m_2 m_2'}(A_i, A_{i'}) C^{l_1 l_2 \lambda}_{m_1' m_2' \mu'} ,
\end{multline}
which is an alternative derivation of Eq. \eqref{eq:wigner-iteration}.
As a final note, we use Eq. \ref{scalar-product} to calculate the $K^{(1)\lambda}_{\mu \mu'}$ kernels. This is very convenient, as most atomistic machine learning software packages can calculate features of the $\langle n \vert \rho_{i'}^{\otimes 1}; \lambda \mu \rangle$ kind, otherwise known as density expansion coefficients in the literature\cite{musi+21cr}.

\section{Equivariance with respect to inversion}\label{app:inversion}

In order to take inversion equivariance into account, we include a further index $\sigma$: $\krn_{\mu \mu'}^{\nu, \lambda \sigma}$. For more details, we redirect the reader to Ref. \cite{niga+20jcp} and its Supplemental Information. Here, it suffices to say that the meaning of the $\sigma$ index with regards to inversion is similar to that of $\lambda$, $\mu$, and $\mu'$ with regards to rotations, and that Eq. \ref{eq:wigner-iteration} needs to be slightly modified as follows:
\begin{multline}\label{eq:equivariant_recursion_with_inversion}
    \krn_{\mu \mu'}^{\nu+1, \lambda \sigma} (A_i, A_{i'}) = \sum_{\substack{l_1 m_1 m_1'l_2 m_2 m_2' \\ \sigma = s_1 s_2 (-1)^{l_1+l_2+\lambda}}}  \\ 
    \krn_{m_1 m_1'}^{\nu, l_1 s_1} (A_i, A_{i'}) \, \krn_{m_2 m_2'}^{1, l_2 s_2} (A_i, A_{i'}) \, C^{l_1l_2\lambda}_{m_1m_2\mu} \, C^{l_1l_2\lambda}_{m_1'm_2'\mu'}.
\end{multline}
The $\nu = 1$ kernels from Eq. \ref{eq:nu1-kernels} always have $\sigma = 1$\cite{musi+21cr}.
We will refer to Eq. \ref{eq:equivariant_recursion_with_inversion}, which calculates the symmetry-adapted kernels with correlation order $\nu + 1$ from those with correlation orders $\nu$ and $1$, as a Wigner iteration.

\section{Scaling analysis} \label{app:scaling}
Since the generation of $\nu = 1$ kernels (Eq. \ref{eq:nu1-kernels} and Appendix \ref{app:features}) is computationally cheap, we will focus on the cost of the Wigner iterations. A simple inspection of Eq. \ref{eq:wigner-iteration} yields the scaling of the computational cost of the algorithm with respect to various convergence parameters.
\paragraph*{$n_{\mathrm{max}}$ and $a_{\mathrm{max}}$:} Since the $n$ index only appears in the initial (and inexpensive) generation of $\nu = 1$ kernels, there is approximately no scaling associated with $n_{\mathrm{max}}$.  The same is true for the element indices $a$, hence the cost of evaluating Wigner kernels is independent of the total number of elements in the system $a_{\mathrm{max}}$. If one wanted to avoid the dependence entirely, it would suffice to compute explicitly the double sum~\eqref{eq:nu1-kernels}, which scales with the product of the number of neighbors in the two environments but does not include any basis.

\paragraph*{$\lambda_{\mathrm{max}}$:}  There are nine angular indices in Eq. \ref{eq:wigner-iteration}: $\lambda$, $\mu$, $\mu'$, $l_1$, $m_1$, $m_1'$, $l_2$, $m_2$, and $m_2'$. While in principle they all scale linearly with $\lambda_{\mathrm{max}}$, $m_2$ and $m_2'$ are redundant due to the properties of the Clebsch-Gordan coefficients. Therefore, the model scales as $\lambda_{\mathrm{max}}^7$. 
\paragraph*{$\nu_{\mathrm{max}}$:} supposing the model is truncated at a correlation order of $\nu_{\mathrm{max}}$ (which will generate a physical model with a maximum body-order of $\nu_{\mathrm{max}} + 1$), $\nu_{\mathrm{max}}-2$ Wigner iterations (Eq. \ref{eq:wigner-iteration}) are needed. If all iterations are truncated at $\lambda \leq \lambda_{\mathrm{max}}$, the cost of the iterations is identical for any $\nu$, and therefore the cost of computing \emph{all} terms scales linearly with $\nu_{\mathrm{max}}$. Note that if a single $\nu$ is desired, the cost can be lowered by performing Wigner iterations between kernels with $\nu>1$, see Appendix~\ref{app:wigner-squares}. More generally, the iteration~\eqref{eq:wigner-iteration} can be trivially generalized to combine arbitrary equivariant kernels, which makes it possible to compute certain forms of symmetrized non-linear equivariant functions more efficiently.

\paragraph*{$n_{\mathrm{train}}$.} 
In a naive implementation of kernel methods, such as what we will use here for simplicity,  training requires computing a $n_\mathrm{train}\times n_\mathrm{train}$ kernel matrix $\mbf{K}$ and inverting it. 
Given the substantial cost of computing Wigner kernel entries, in almost all cases we consider the cost is dominated by the quadratic scaling in computing the kernels, and not by the cubic cost of the inversion. 
Inference has a cost scaling linearly with $n_\mathrm{train}$. 
Most practical KRR implementations use low-rank approximations of the kernel matrix (as in the projected process approximation\cite{rasm06book}) that make the construction of the training matrix formally linear in $n_\text{train}$, and inference independent of it -- similar to methods based on linear regression such as ACE or MTP. 
In practice, however, increasing the sparse set (or feature set) size is inevitable in practice in order to avoid saturation of the accuracy as $n_\mathrm{train}$ increases \cite{bigi+22jcp}.

\section{Practical implementation}
\label{app:implementation}

KRR models such as those discussed in Sec.~\ref{sec:krr} are fully defined by the choice of the kernel function, which therefore strongly affects the accuracy they can achieve.
In our construction, there are two main steps which influence the final kernel function: the definition of the atomic densities and the mixing of different body-ordered kernels.

\subsubsection{Density expansion form}

The general form of the density expansion has already been introduced in Sec. \ref{sec:densities-kernels}. Contrary to common practice, we allow the width of the Gaussians in Eq. \ref{eq:density-expansion} to vary with the distance from the central atom. In particular, we use normalized Gaussians where the width increases exponentially with the distance,
\begin{equation}\label{eq:gaussian-form}
 g(\bx - \br_{ji}) \sim  \exp[-(\bx- \br_{ji})^2/2(C e^{r_{ji}/r_0})^2].
\end{equation}
The cutoff function is set to
\begin{equation}
    f_{\mathrm{cut}}(r_{ji}) = e^{-r_{ji}/r_0},
\end{equation}
and the densities are further multiplied by a shifted cosine function in the last $0.5$ \AA \ before $r_\mathrm{cut}$ to ensure that the predictions of the model (as well as their first derivatives) are smooth as neighbors enter and leave the atomic environments defined by the cutoff radius.
This choice of Gaussian smearing and cutoff function assumes the exponential decrease of the magnitude and resolution of physical interactions with distance.
The hyperparameters $C$ and $r_0$, that determine the the maximum resolution and the decay length, are optimized separately for each dataset via a grid search.

\subsubsection{Exponential kernels}\label{sec:exp-kernels}

Similar to how a modulation of the neighbor density can be used to exploit the physical prior that interactions between atoms decay with distance, one can incorporate the common wisdom that low-order correlations dominate the contributions to atom-centered properties by building a the overall kernel as a 
linear combination of the body-ordered Wigner kernels:
\begin{equation}\label{eq:kernel-lc}
    \krn_{\mu \mu'}^{\lambda \sigma} = \sum_{\nu = 0}^{\nu_\mathrm{max}} c_\nu \krn_{\mu \mu'}^{\nu, \lambda \sigma}.
\end{equation}
While in principle all the $c_\nu$ could vary independently, we found an exponential-like parametrization of Eq. \ref{eq:kernel-lc} to be particularly effective:
\begin{equation}
    \krn_{\mu \mu'}^{\lambda \sigma} = c_0 \krn_{\mu \mu'}^{0, \lambda \sigma} + \, a \sum_{\nu = 1}^{\nu_\mathrm{max}} \frac{b^\nu}{\nu!} \krn_{\mu \mu'}^{\nu, \lambda \sigma},
    \label{eq:exp-kernel}
\end{equation}
A similar kernel construction was proposed in Ref. \cite{glie+18prb} for body-ordered invariant kenels, noting however that no practical algorithm existed for its evaluation.

In this context, the Wigner iteration provides an efficient way to evaluate a truncated Taylor expansion of the exponential by pre-computing the body-ordered kernels $\krn_{\mu \mu'}^{\nu, \lambda \sigma}$ up to $\nu=\nu_\mathrm{max}$. Furthermore, given that it can be applied to combine any pair of equivariant kernels, it would also allow to implement other, more efficient algorithms to evaluate an exponential\cite{mole-vanl03siamr}, such as scaling and squaring (see Appendix~\ref{app:wigner-squares}). In practice, however, we found very high-body-order interactions to be of marginal importance in our tests, so we prefer to evaluate the summation explicitly, as this also simplifies the optimization of the related hyperparameters. These are $c_0$, $a$, and $b$, and they are optimized by dual annealing using 10-fold cross-validation within the training set.  Given that these three coefficients also set the overall scale of the kernel, the regularization that appears in kernel ridge regression is redundant, and we keep it constant.

\section{Generalized Wigner iterations}
\label{app:wigner-squares}

Even though one cannot compute element-wise non-linear functions of kernels with $\lambda\ne 0$ without disrupting their equivariant behavior, it is possible to define equivariant non-linear functions of the kernels through their Taylor expansion, e.g.,
\begin{equation}
\exp(\krn^{\lambda\sigma}_{\mu\mu'}) \equiv
\sum_{n=0}^\infty 
\frac{1}{n!} \krn^{n, \lambda\sigma}_{\mu\mu'}.
\end{equation}
Much as it is the case for matrix functions, one can apply several tricks to evaluate these quantities more efficiently than through a truncated series expansion. For example, one can evaluate the exponential through a scaling-and-squaring relation
\begin{equation}
\exp(\krn^{\lambda\sigma}_{\mu\mu'}) =
[\exp(\krn^{\lambda\sigma}_{\mu\mu'}/2^p)]^{2^p}.
\end{equation}
One first computes $\exp(\krn^{\lambda\sigma}_{\mu\mu'}/2^p)$ with a low-order expansion (which works because $2^p$ makes the argument of the exponential very small) and then apply the generalized Wigner iteration $p$ times, multiplying each time the result by itself. 

In a similar spirit, if one is only interested in the calculation of all \emph{invariant} kernels up to $\nu = \nu_\text{max}$, the Wigner iteration procedure can be simplified. Indeed, it is sufficient to perform full (equivariant) Wigner iterations only up to $\lceil \nu_\text{max}/2 \rceil$ and then combine low-order equivariant kernels to get high-order invariant kernels. For example, if $\nu_\text{max}$ is even, $\krn^{\nu_\text{max},01}_{00}$ can be calculated as the product of the $\krn^{\nu_\text{max}/2,\lambda\sigma}_{\mu \mu'}$ kernels with themselves via an inexpensive invariant-only Wigner iteration. Due to the $\lambda \leq \lambda_\text{max}$ truncation strategy, these kernels might not exactly correspond to those calculated via full Wigner iterations. However, we did not find any differences in performance between the two evaluation strategies, which simply correspond to slightly different angular truncations of the high-order kernels.

\section{Hyperparameters}\label{app:hypers}
\label{app:fit-details}

\subsubsection{Wigner kernels}
When it comes to the usability of a model, a distinction should be made between ``convergence'' hyperparameters and ``optimizable'' hyperparameters. The former are those that show a monotonic improvement of the accuracy of the model as they are increased, but which need to be set to a finite value for practical feasibility. The question then becomes whether they can be converged without compromising the computational speed of the model. In the Wigner kernel case, these are $r_\text{cut}$, $\nu_\text{max}$, $\lambda_\text{max}$, and $n_\text{max}$.

\begin{itemize}
    \item $r_\text{cut}$ only enters the initial calculation of the $\nu = 1$ kernels. As a result, the model's training and evaluation times are virtually unaffected by its choice, as long as it is not too large.
    \item The same is true for $n_\text{max}$, i.e., the number of radial basis functions used to calculate the $\nu = 1$ kernels: as it does not enter the Wigner iterations, it can be converged almost arbitrarily. Using a Laplacian eigenstate radial basis\cite{bigi+22jcp}, we did not notice any significant improvement to the accuracy of the models past $n_\text{max} = 25$, hence we set it to that value for all benchmarks.
    \item In contrast, the number of Wigner iterations needed to evaluate the kernels grows linearly with $\nu_\text{max}$. We did not find $\nu_\text{max}$ to limit the accuracy or the computational cost of the Wigner kernel model in any of our benchmarks.
    \item $\lambda_\text{max}$ is the most critical of these convergence hyperparameters, as the computational performance of the proposed model depends heavily on it. Although going past $\lambda_\text{max} = 3$ or $4$ is impractical with our current implementation, our results do not identify this limitation as critical to improve the accuracy of the model. We provide a tentative explanation of this phenomenon in Section \ref{sec:methane}.
\end{itemize}
The convergence hyperparameters used in this work are reported in Table~\ref{tab:convergence-hypers}.
  
In addition, the Wigner kernel model as presented in this work has two \emph{optimizable} hyperparameters. These are the $C$ and $r_0$ coefficients that enter the density definition in Eq. \ref{eq:gaussian-form}. Given their very small number, we optimize these via a grid search. The fact that, in the formulation we present here, two physically-interpretable density modulation parameters determine the value of the kernel is a significant advantage of our framework. 
As discussed in the main text, the exponential-like kernel parameters can be optimized without having to re-compute the kernels, and we optimize them automatically by cross-validation within the training set.  %

\begin{table}[h]
    \centering
    \begin{tabular}{cccc}
    \hline\hline
        Model & $\rcut$ (\AA) & $\nu_\text{max}$ & $\lambda_\text{max}$ \\ \hline
        Methane  & 6.0 & 4 & 0, 1, 2, 3 \\
        Gold  & 6.0 & 2, 3, 4, 6 & 3 \\
        QM9  & 5.0 & 4 & 3 \\
    \hline\hline
    \end{tabular}
    \caption{``Convergence'' hyperparameters used for the WK models in Sec.~\ref{sec:results}.}
    \label{tab:convergence-hypers}
\end{table}

\subsubsection{SOAP-GPR and linear SOAP}

In our benchmarks, we also provide fits for the SOAP-GPR and linear SOAP methods. The SOAP descriptors\cite{bart+13prb} present a large number of hyperparameters. In the case of SOAP-GPR, these need to be added to the choice of the kernel\cite{deri+21cr}. With such a large hyperparameter space, it is almost mandatory to rely on previous knowledge and common practice. 
Hence, for the random methane dataset, we employed a GTO basis with $l_\text{max} = 6$, $n_\text{max} = 8$, a Gaussian smearing of 0.2 \AA \ which was found to be optimal in Ref.~\citenum{pozd+21ore}, and the same radial scaling that was used for the QM9 dataset in Ref.~\citenum{will+18pccp}. 
For the gold cluster dataset, we used the same SOAP hyperparameters that were used in the silicon fit in Ref.~\citenum{gosc+21jcp}. A squared kernel was used in all SOAP-GPR models, as it is one of the most common choices\cite{deri+21cr}.

\subsubsection{LE-ACE}

In this work, LE-ACE was benchmarked on the gold cluster dataset. The LE-ACE model as presented in Ref.~\cite{bigi+22jcp} has $\nu_\text{max}+1$ hyperparameters, roughly corresponding to the maximum Laplacian eigenvalues for $\nu = 1, ..., \nu_\text{max}$ plus a radial transform parameter. Since we used $\nu_\text{max} = 6$, we found the resulting hyperparameter space to be impossible to optimize rigorously and we therefore optimized it heuristically. We suspect that the successful use of exponential kernels in this work will provide valuable insights in designing more compact and effective hyperparameter spaces for models such as ACE.

\section{Angular scans and kernel resolution}
\label{app:angular-scans}

\begin{figure}
    \centering
    \includegraphics[width=\linewidth]{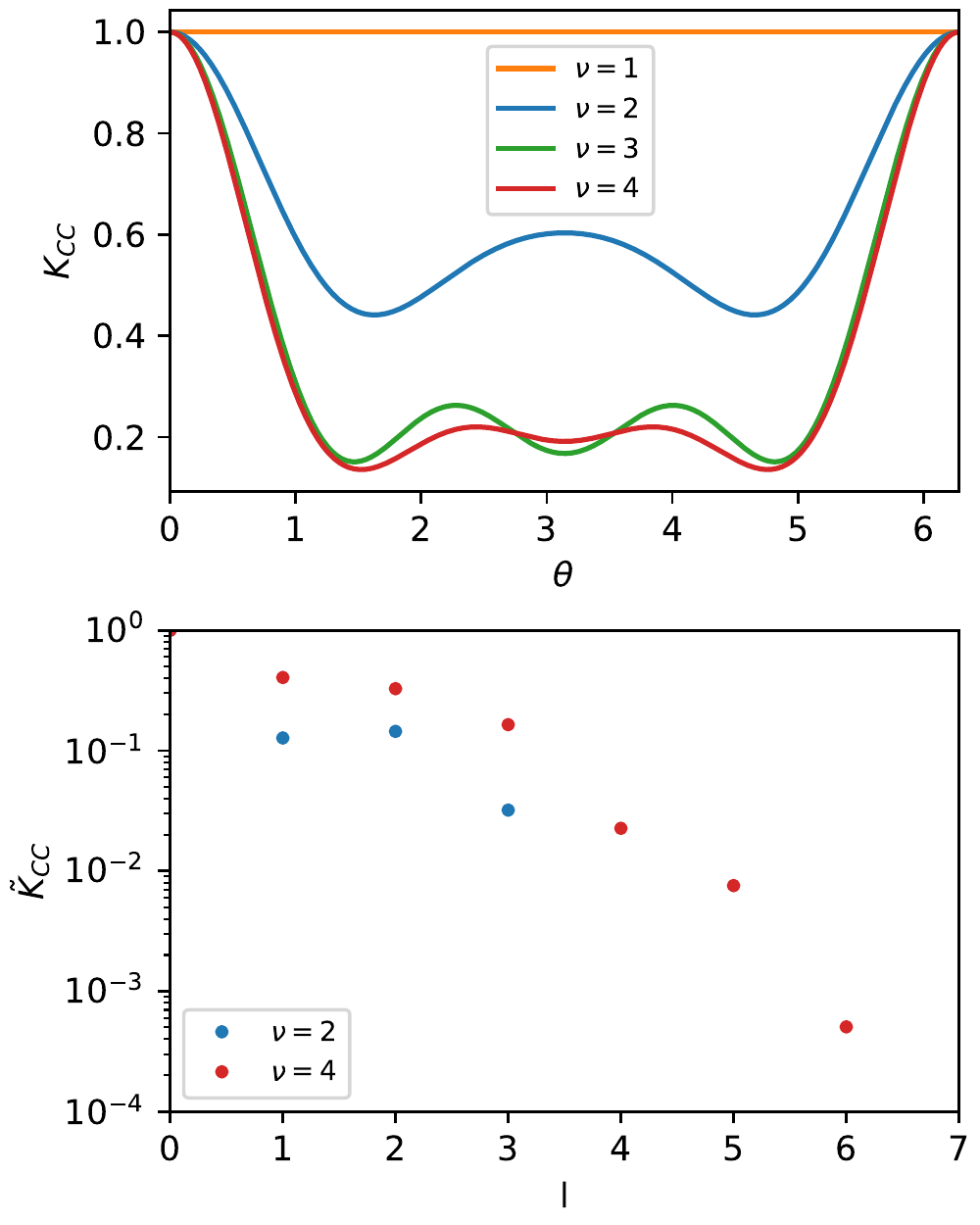}
    \caption{Top panel: angular scan showing a carbon-centered kernel between a \ce{CH4} molecule and a \ce{CH2} molecule. The \ce{CH4} molecule is a random molecule from the methane dataset, while the \ce{CH2} molecule has C-H distances of 1.0 and 1.5 \AA, and the H-C-H angle $\theta$ is free to rotate. Bottom panel: Fourier transform coefficients of the curves in the top panel at frequencies $l/2\pi$, showing that, although $\lambda_\text{max} = 3$ for both kernels, the $\nu = 4$ kernel contains higher ($l>3$) frequency components. }
    \label{fig:scan}
\end{figure}

To demonstrate the increase in resolution afforded by high-$\nu$ kernels, we compute \ce{C}-centered Wigner kernels between a random \ce{CH4} environments and a set of \ce{CH2} structures where we vary the \ce{H-C-H} angle ($\theta$) for fixed \ce{C-H} distances. 
This experiment reveals how higher-$\nu$ kernels are capable of describing higher-frequency components of the \ce{H-C-H} angular correlations (Fig.~\ref{fig:scan}). Thus, body-order and structure-space resolution are not fully decoupled. 

\end{document}